\pdfoutput=1

\documentclass[entropy,article,accept,moreauthors,pdftex]{Definitions/mdpi}
\firstpage{1}
\pubvolume{23}
\issuenum{6}
\articlenumber{284}
\pubyear{2021}
\copyrightyear{2021}
\externaleditor{Academic Editors: Antonino Messina and Agostino Migliore} 
\datereceived{30 April 2021}
\dateaccepted{25 May 2021}
\datepublished{28 May 2021}
\hreflink{https://doi.org/10.3390/e23060684} 

\Title{Semi-Classical Discretization and Long-Time Evolution of Variable Spin
Systems}

\TitleCitation{Semi-Classical Discretization and Long-Time Evolution of Variable Spin Systems}

\Author{Giovani E. Morales-Hern\'andez, Juan C. Castellanos, Jos\'e L. Romero\orcidA{} and Andrei B. Klimov *\orcidB{}}

\AuthorNames{Giovani E. Morales-Hernández, Juan C. Castellanos, José L. Romero and Andrei B. Klimov}

\AuthorCitation{Morales-Hern\'andez, G.E.; Castellanos, J.C.; Romero, J.L.; Klimov, A.B.}

\address[1]{%
Departamento de F\'isica, Universidad de Guadalajara, Guadalajara 44420, Jalisco, Mexico; giovani.morales8917@alumnos.udg.mx (G.E.M.-H.); juan.castellanos@alumnos.udg.mx (J.C.C.); jose.romero@cucei.udg.mx (J.L.R.)\\
\corres{\hangafter=1 \hangindent=1.05em \hspace{-0.82em}  Correspondence: klimov@cencar.udg.mx}}

\abstract{We apply the semi-classical limit of the generalized $SO(3)$ map for
representation of variable-spin systems in a four-dimensional symplectic
manifold and approximate their evolution terms of effective classical
dynamics on $T^{\ast }\mathcal{S}_{2}$. Using the asymptotic form of the
star-product, we manage to ``quantize'' one of the classical dynamic variables and
introduce a discretized version of the Truncated Wigner Approximation (TWA).
Two emblematic examples of quantum dynamics (rotor in an external field and
two coupled spins) are analyzed, and the results of exact, continuous, and
discretized versions of TWA are compared.}



\begin{document}
\section{Introduction}

Phase-space methods provide a very convenient framework for the analysis of
large quantum systems (QS) \cite{Schroeck(1996),
Schleich(2001),Zachos(2005),Almeida(1998)}. In cases when the states of QS
are elements of a Hilbert space $\mathbb{H}$ which carries a unitary
irreducible representation of a Lie group $G$, trace-like maps from
operators $\hat{f}$ \ acting in $\mathbb{H}$ into their (Weyl) symbols $%
W_{f}(\Omega )$ can be established. The symbols $W_{f}(\Omega )$ are
functions on the corresponding classical phase-space $\mathcal{M}$, $\Omega
\in \mathcal{M}$ being the phase-space coordinates. The most suitable in
applications is frequently the Wigner (self-dual) map, allowing  the
average value of an observable to be computed as a convolution of its
symbol with the symbol of the density matrix $W_{\rho }(\Omega )$ (the
Wigner function). The properties of the Wigner function are extremely useful
for studying the quantum-classical correspondence. In particular, the
quantum evolution is described by a partial differential equation (the Moyal
equation \cite{Moyal(1949)}) for the Wigner function with a well-defined
classical limit. In the non-harmonic case, the Moyal equation
contains high-order derivatives, which turns the finding of its exact
solution into quite a difficult task.

The advantage of the phase-space approach to quantum dynamics
consists of the possibility of expanding the Moyal equation in powers of
small parameters in the semi-classical limit. Such semi-classical
parameters are related both to the symmetry of the Hamiltonian and of the
map.

In general, different mappings can be performed for composite
quantum systems. The type of map fixes the structure of the phase-space
manifold, and thus the allowed set of shape-preserving transformations. In
addition, the phase-space symmetry determines the physical semiclassical
parameter $\varepsilon \ll 1$.

The standard phase-space approach \cite{Perelomov(1986), Zhang(1990),
Gadella(1995), Brif(1999), Klimov(2009)} is not applicable for the
construction of covariant (under group transformations) invertible maps if
the density matrix cannot be decomposed into a direct sum of components each
acting in an irreducible representation of the dynamical group $G$.
Physically, this happens when the Hamiltonian of the system induces
transitions between irreducible subspaces i.e., the total angular
momentum is changed in the course of evolution. This happens, for
instance, in quantum systems with non-fixed (variable) spin, such as large
interacting spins, a rigid rotor in an external field, coupled and
externally pumped boson modes with and without decay, etc.

A suitable $SU(2)$ covariant map, establishing a one-to-one relation
between operators and set of functions (discretely labelled symbols) in a
three-dimensional space, can be found for variable spin systems \cite%
{Klimov(2008)}. In addition, there exists a continuous limit of such symbols
for large values of the mean spin \cite{Tomatani(2015)}. This allows the
quantum operators to be put in correspondence with smooth functions on the
four-dimensional cotangent bundle $T^{\ast }\mathcal{S}_{2}$, equipped with
a symplectic structure. In particular, the semi-classical dynamics of the
Wigner function of variable spin systems can be described in terms of
effective \textquotedblleft classical\textquotedblright\ trajectories $%
\Omega ^{cl}(t)$ in the phase-space $T^{\ast }\mathcal{S}_{2}$. As a rough
approach, the evolution of average values can then be estimated within the
framework of the so-called Truncated Wigner Approximation (TWA), which
consists of propagating points of the initial distribution along the
classical trajectories. Such an approximation has been successfully applied
for studying a short-time evolution of QS with different dynamic symmetry
groups \cite{Heller(1976), Heller(1977), Davis(1984), Kinsler(1993),
Drobny(1992), Drobny(1997), Polkovnikov(2010), Amiet(1991), Klimov(2002),
Klimov(2005), Kalmykov(2016), de_Aguiar(2010), Viscondi(2011),
Gottwald(2018), Klimov(2017)}. Expectably, the TWA fails to describe the
non-harmonic evolution beyond the Ehrenfest (or semiclassical) time $\tau
_{sem}$, \cite{Ehrenfest(1927), Zaslavsky(1981), Hagedorn(2000),
Silvestrov(2002), Schubert(2012)}, where intrinsic quantum correlations
effects start to emerge. The semi-classical time heavily depends both on the
Hamiltonian and on the initial state (in particular on the stability of the
classical motion). For instance, for spin $S$ systems, the semiclassical
time usually scales as the inverse power of the spin length, $gt\lesssim
S^{-\alpha }$, $\alpha >0$, where $g$ is a constant characterizing the
non-harmonic dynamics. The semiclassical time for variable spin systems
behaves in a similar way, where the effective spin size is proportional to
the average total angular momentum.

Multiple attempts to improve the TWA \cite{Filinov(2008), Schubert(2009),
Polkovnikov(2010)} suggest that more sophisticated methods \cite%
{Dittrich(2006), Dittrich(2010), Maia(2008), Toscano(2009), Almeida(2013),
Tomsovic(2018), Lando(2019), Rios(2002), Ozorio(2006)} should be applied in
order to describe the quantum evolution in terms of continuously distributed
classical phase-space trajectories. Alternatively, different types of
discrete phase-space sampling were proposed \cite{Takahashi(1993),
Schachenmayer_Many(2015), Schachenmayer_Dynamics(2015), Acevedo(2017),
Pineiro(2017), Pucci(2016), Sundar(2019)} in order to emulate evolution of
average values using the main idea of the TWA. In general, phase-space
discretization is a tricky question, which has been addressed from different
perspectives mainly focusing on the flat and torus phase-space manifolds
\cite{Littlejohn(2002), Light(2000)}.

It is worth noting that the phase-space analysis of some variable spin
systems, such as, for example, two large interacting spins, can be formally
performed using the Schwinger representation. Then, faithful mapping onto a
flat $C^{2}\otimes C^{2}$\ phase-space is carried out by applying the
standard Heisenberg-Weyl $H(1)\times H(1)$\ map \cite%
{Glauber(1963),Sudarshan(1963),Cahill(1969)1857,Cahill(1969)1882}. However,
the natural $SU(2)$ symmetry is largely lost in such an approach. This is
reflected, in the fact that (a) the corresponding distributions are not
covariant under rotations; and (b) the inverse excitation numbers in each of
the boson modes play the role of \textit{dynamical} semi-classical
parameters. The explicit time-dependence of such semi-classical parameters
may restrict the validity of the formal division of the Moyal equation to
the classical part, containing only the Poisson brackets and the so-called
``quantum corrections''. All of this leads to inefficiency of the standard
semiclassical methods \cite{Polkovnikov(2010)}. The map of the same systems
onto the $\mathcal{S}_{2}\otimes \mathcal{S}_{2} $-spheres in the
Stratonovich--Weyl framework \cite{Stratonovich(1957), Agarwal(1981),
Varilly(1989)} reveals only the local $SU(2)$ symmetry. The semiclassical
parameters are the inverse spin lengths (constant in time). Thus, the $SU(2)$
TWA leads, in principle, to better results than its flat counterpart in $%
C^{2}\otimes C^{2}$. However, the standard discretization of the $\mathcal{S}%
_{2}$ sphere \cite{Sun(2008)} does not lead to a significant improvement of
the TWA, since the location of the initial distribution is not taken into
account.

The situation is even more intricate when the number of involved invariant
subspaces becomes formally infinite (or physically very large), as, for
example, in the case of a highly-excited rigid rotor interacting with
external fields. The use of the Schwinger representation leads to
substantial complications in both the analytical and numerical calculations
and the standard $SU(2)$ map is simply non-applicable.
Thus, the analysis of the semi-classical limit becomes very challenging \cite%
{Harter(1984),Schmiedt(2017)}.

In the present paper, we show that there is a natural discretization of $%
T^{\ast }\mathcal{S}_{2}$ in the vicinity of the initial distribution, which
allows the time-scale of validity of the TWA to be significantly increased,
including the so-called revival times. Such a discretization is based on
the asymptotic form of the star-product for spin-variable systems \cite%
{Tomatani(2015)} and is directly applicable to calculations of the evolution
of mean values of physical observables. Using our approach, we will be able
to describe the long-time dynamics of molecule in an external field
(modelled by a physical rotor) and a coupled two-spin system in the
semi-classical limit. We restrict our study to quantum systems with $SO(3)$
symmetry, corresponding to integer spins.

The paper is organized as follows: In Section \ref{basis}, we recall the
basic results on the $SO(3)$ covariant mapping for variable spin systems. In
Section~\ref{semiclassical limit}, we discuss the asymptotic form of the
star-product in the semi-classical limit. In Section~\ref{discretization},
we develop a discretization scheme on the classical manifold of variable
spin systems and apply it to computation of mean values in a ``quantized''
version of the TWA. Two applications of the proposed method with the
corresponding numerical solutions are discussed in Section~\ref{examples}. A
summary and conclusions are given in Section~\ref{conclusions}.


\section{Variable Spin Wigner Function}

\label{basis}

Let us consider a QS whose states are elements of a Hilbert space $\mathbb{H}
$ containing multiple $SO(3)$ irreps, so that the $SO(3)$ group, in general, does
not act irreducibly on the density matrix of the system $\hat{\rho}$, i.e.,%
\[
\hat{\rho}=\sum_{L,L^{\prime }=0,1,..}\sum_{m,m^{\prime }}c_{mm^{\prime
}}^{LL^{\prime }}|L,m\rangle \langle L^{\prime },m^{\prime }|. 
\]

The generalized Wigner-like map \cite{Klimov(2008)} from operators acting
in $\mathbb{H}$ to a discrete set of functions, later called $j$-symbols,
\begin{equation}
\hat{f}\Leftrightarrow \{W_{f}^{j}\left( \Theta \right) ;\,j=0,1,\dots\},
\label{fW}
\end{equation}%
where
\begin{equation}
\Theta =(\phi ,\theta ,\psi )\in \mathcal{S}_{3},\,0\leq \phi <2\pi,\,0\leq
\theta <\pi ,\,0\leq \psi <2\pi  \label{Theta}
\end{equation}
are the Euler angles, is defined through a trace operation

\begin{equation}
W_{f}^{j\ }\left( \Theta \right) =\mathrm{Tr}\left( \hat{f}\hat{\omega}%
_{j}\left( \Theta \right) \right) ,  \label{j_sym}
\end{equation}%
where the Hermitian $SO(3)$ covariant mapping kernels have the form%
\begin{equation}
\hat{\omega}_{j}\left( \Theta \right) =\sum_{K=0}^{j}\sum_{Q,Q^{\prime
}=-K}^{K}\sqrt{\frac{2K+1}{j+1}}D_{QQ^{\prime }}^{K}\left( \Theta \right)
\hat{T}_{KQ}^{\frac{j+Q^{\prime }}{2}\ \frac{j-Q^{\prime }}{2}},
\label{kernel}
\end{equation}%
where $D_{QQ^{\prime }}^{K}\left( \Theta \right) $ is the Wigner $D$%
-function, $D_{QQ^{\prime }}^{K}\left( \Theta \right) =\langle K,Q|e^{-i\phi 
\hat{l}_{z}}e^{-i\theta \hat{l}_{y}}e^{-i\psi \hat{l}_{z}}|K,Q^{\prime
}\rangle $, here $\hat{l}_{x,y,z}$ are generators of $SO(3)$ group, $[\hat{l}%
_{k},\hat{l}_{m}]=i\varepsilon _{kmn}\hat{l}_{n}$, 
\begin{equation}
\hat{T}_{KQ}^{JJ^{\prime }}=\sum_{M,M^{\prime }}\sqrt{\frac{2K+1}{2J+1}}%
C_{J^{\prime }M^{\prime },\ KQ}^{JM}\left\vert J,M\right\rangle \left\langle
J^{\prime },M^{\prime }\right\vert ,  \label{ITO}
\end{equation}%
are tensor operators \cite{Blum(2012)} and $C_{a\alpha ,\ b\beta }^{c\gamma
} $ are the Clebsch--Gordan coefficients. The map~\eqref{j_sym} is
explicitly invertible
\begin{eqnarray}
\hat{f} &=&\sum_{j=0,1\dots }^{\infty }\hat{f}_{j},  \label{recon2} \\
\hat{f}_{j} &=&\frac{j+1}{8\pi ^{2}}\int d\Theta W_{f}^{j}\left( \Theta
\right) \hat{\omega}_{j}\left( \Theta \right),
\end{eqnarray}%
where $d\Theta =\sin \theta d\phi d\theta d\psi $ is a volume element of $%
SO\left( 3\right) $, leading to the overlap relation
\begin{equation}
\mathrm{Tr}\left( \hat{f}\hat{g}\right) =\sum_{j=0,1,2,\dots }^{\infty }%
\frac{j+1}{8\pi ^{2}}\int d\Theta W_{f}^{j}\left( \Theta \right)
W_{g}^{j}\left( \Theta \right).  \label{over}
\end{equation}

It is worth noting that the operators $\hat{f}_{j}$ correspond to the
expansion of $\hat{f}$ on the tensor operators $\hat{T}_{KQ}^{JJ^{\prime }}$~%
\eqref{ITO} in the sectors with fixed values of $j=J^{\prime }+J$:
\begin{eqnarray}
\hat{f}_{j} &=&\sum_{K=0}^{j}\sum_{Q,Q^{\prime }=-K}^{K}\hat{T}_{K\,Q}^{%
\frac{j+Q^{\prime }}{2}\,\frac{j-Q^{\prime }}{2}}f_{K\,Q}^{\frac{j+Q^{\prime
}}{2}\,\frac{j-Q^{\prime }}{2}},  \label{fj} \\
f_{K\,Q}^{\frac{j+Q^{\prime }}{2}\,\frac{j-Q^{\prime }}{2}} &=&Tr\left( \hat{%
f}\hat{T}_{K\,Q}^{\frac{j+Q^{\prime }}{2}\,\frac{j-Q^{\prime }}{2}\dagger
}\right).
\end{eqnarray}

It should be noted that $Q^{\prime }$ in~\eqref{kernel}, running over even or
odd integers depending on the parity of the index $j$, such that the
restriction $Q^{\prime }\pm j$ is an even number, is fulfilled. This leads
to the following symmetry properties of the kernel:%
\begin{eqnarray}
\hat{\omega}_{j}(\phi ,\theta ,\psi ) &=&\hat{\omega}_{j}(\phi ,\theta ,\psi
+\pi ),\text{ even }j\text{,}  \label{we} \\
\hat{\omega}_{j}(\phi ,\theta ,\psi ) &=&-\hat{\omega}_{j}(\phi ,\theta
,\psi +\pi ),\text{ odd }j\text{.}  \label{wo}
\end{eqnarray}

The advantage of the generalized map~\eqref{fW}--\eqref{j_sym} consists of
the possibility of a ``classical'' representation of the whole operator acting
in $\mathbb{H}$ and not only its projections on the $SO(3)$ irreducible
subspaces. For instance, for the orientation operators, $\mathbf{\hat{r}}$,
\begin{eqnarray}
\mathbf{\hat{r}} &=&\int d\varphi \,\sin \vartheta d\vartheta \,\mathbf{n(}%
\varphi ,\vartheta )|\varphi ,\vartheta \rangle \langle \varphi ,\vartheta
|,\quad \quad \mathbf{\hat{r}}^{2}=\hat{I},  \label{nz0} \\
|\varphi ,\vartheta \rangle &=&\sum_{J=0,1,\dots
}\sum_{M=-J}^{J}Y_{JM}^{\ast }(\varphi ,\vartheta )|J,M\rangle ,
\label{x base}
\end{eqnarray}%
where $\varphi $, $\vartheta $ are angles in the
configuration space, $\mathbf{n}(\varphi ,\vartheta )=\left( \cos \varphi
\sin \vartheta ,\sin \varphi \sin \vartheta ,\cos \vartheta \right) $,
one obtains
\begin{equation}
W_{r_{k}}^{j}(\Theta )=r_{k}\sum_{n\in \mathbb{Z}^{+}}\delta _{j,2n+1},
\label{Wr}
\end{equation}%
{where $\mathbf{r}=\left( \sin {\phi }\sin {\psi }-\cos {\phi }\cos {\theta }%
\cos {\psi },-\cos {\phi }\sin \psi -\sin {\phi }\cos {\theta }\cos {\psi }%
,\sin {\theta }\cos {\psi }\right) $, \linebreak\mbox{$\mathbf{r}^{2}=1$.}}
The dependence of the symbol on the angle $\psi $ indicates that the
corresponding operator mixes $SO(3)$ invariant subspaces. Vice-versa,
symbols of the operators that preserve each $SO(3)$ irreducible subspace are
independent of ${\psi }$, as, for instance, the angular momentum operators, $%
\mathbf{\hat{l}}=(\hat{l}_x,\hat{l}_y,\hat{l}_z )$,%
\begin{equation}
W_{l_{k}}^{j}(\Theta )=\sqrt{\frac{j}{2}\left( \frac{j}{2}+1\right) }%
\,n_{k}(\phi ,\theta )\sum_{n\in \mathbb{Z}^{+}}\delta _{j,2n},  \label{Wl}
\end{equation}%
where $\mathbf{n}(\phi ,\theta )=\left( \cos \phi \sin \theta ,\sin \phi
\sin \theta ,\cos \theta \right) $ is a unitary vector in the parameter
space~\eqref{Theta}, and, as a consequence,%
\begin{equation}
W_{\mathbf{l}^{2}}^{j}\!(\Theta )=\frac{j}{2}\left( \frac{j}{2}+1\right)
\sum_{n\in \mathbb{Z}^{+}}\delta _{j,2n},  \label{l2}
\end{equation}%
where $\mathbf{\hat{l}}^{2}$ is the square angular momentum operator.

The standard Stratonovich--Weyl kernel $\hat{w}_{L}(\phi ,\theta )$ \cite%
{Stratonovich(1957), Agarwal(1981), Varilly(1989)}, used for mapping
operators acting in a single $SO(3)$ subspace of dimension $2L+1=j+1$ ($L$ \
is an integer),%
\begin{equation}
\hat{f}\Leftrightarrow W_{f}\left( \phi ,\theta \right) =Tr\left( \hat{f}%
\hat{w}_{L=j/2}(\phi ,\theta )\right) ,  \label{su2map}
\end{equation}%
is recovered from the generalized kernel $\hat{\omega}_{j}(\Theta )$~%
\eqref{kernel} by integrating over the angle $\psi $ (for even values of $j$%
): 
\begin{equation}
\hat{w}_{L=j/2}(\phi ,\theta )=\int_{0}^{2\pi }\frac{d\psi }{2\pi }\,\hat{%
\omega}_{j}(\Theta )=\sqrt{\frac{4\pi }{2L+1}}\sum_{K=0}^{2L}\sum_{Q=-K}^{K}%
\,Y_{KQ}^{\ast }(\phi ,\theta )\,\hat{T}_{KQ}^{L}\,  \label{wsu2}
\end{equation}%
where $Y_{KQ}(\phi ,\theta )$ are spherical harmonics and $\hat{T}_{KQ}^{L}$
are the standard (diagonal) tensor operators \cite{Varshalovich(1988),
Biedenharn(1984)}.

It is important to stress that the kernel (\ref{kernel}) is not reduced to
the direct product of the standard $SO(3)$ kernels \eqref{wsu2}. Therefore,
the map (\ref{fW}), possessing the underlying global $SO(3)$ symmetry,
allows us to faithfully represent operators in the form of $c$-functions
that:

(a) act in two independent $SO(3)$ irreps, as, for example, a direct product
of angular momentum operators $\hat{l}_{k}^{(1)}\otimes \hat{l}_{m}^{(2)}$.
It should be observed that an alternative mapping can also be achieved with
the kernel $\hat{w}_{L_{1}}(\phi ,\theta )\otimes \hat{w}_{L_{2}}(\phi
,\theta )$. However, in the latter case, the underlying symmetry group is $%
SO(3)\times SO(3)$. The advantage of one of the map over another is not
obvious. It will be shown below that the map (\ref{fW}) admits a natural
discretization in the semiclassical limit that significantly improves the
range of applicability of the Truncated Wigner Approximation;

(b) mixes all $SO(3)$ irreps, as, for example, the orientation operator (\ref%
{nz0}). This type of operators cannot be mapped into their classical
counterparts in the framework of the standard Stratonovich--Weyl approach (%
\ref{su2map}).

\section{Wigner Function Dynamics in the Semi-Classical Limit}

\label{semiclassical limit}

The crucial feature of the map~\eqref{fW}--\eqref{j_sym} is the possibility
of introducing a star-product operator \cite{Moyal(1949), Bayen(1978)},
acting on $j$-symbols \cite{Klimov(2008)}:%
\begin{equation}
W_{fg}^{j}(\Theta )=\sum_{j_{1},j_{2}=0,1,\dots}L^{j,j_{1}j_{2}}\left(
W_{f}^{j_{1}}\left( \Theta \right) W_{g}^{j_{2}}\left( \Theta \right)
\right) ,  \label{star}
\end{equation}%
which is reduced to the standard (local) form \cite{Klimov_Espinoza(2002),
Rios(2014)} when the operators $\hat{f}$ and $\hat{g}$ are operators from
the $se(3)$ enveloping algebra. The exact form of the star-product operator
in general is non-local on the index $j$ and has an involved form, but it
is significantly simplified in the limit $j\gg 1$ \cite{Tomatani(2015),
Klimov(2017)},%
\begin{eqnarray}
L^{j,j_{1}j_{2}} &\approx &\mathcal{V}\int_{0}^{2\pi }\frac{d\varphi
d\varphi ^{\prime }}{\left( 2\pi \right) ^{2}}e^{i\left( j_{2}-j+\mathbb{J}%
^{0}\otimes I\right) \varphi }e^{i\left( j_{1}-j-I\otimes \mathbb{J}%
^{0}\right) \varphi ^{\prime }},  \label{exp} \\
\mathcal{V} &=&e^{-\varepsilon \mathbb{J}^{0}\otimes \mathbb{J}^{0}-\frac{%
\varepsilon }{2}\left( \mathbb{J}^{+}\otimes \mathbb{J}^{-}-\mathbb{J}%
^{-}\otimes \mathbb{J}^{+}\right) },  \label{V}
\end{eqnarray}%
where $\varepsilon =(j+1)^{-1}$,
\begin{equation}
\mathbb{J}^{\pm }=ie^{\mp i\psi }\left[ i\frac{\partial }{\partial \theta }%
\pm \cot \theta \frac{\partial }{\partial \psi }\mp \frac{1}{\sin \theta }%
\frac{\partial }{\partial \phi }\right] \,\quad \mathbb{J}^{0}=-i\frac{%
\partial }{\partial \psi }\,  \label{Jops}
\end{equation}%
and the notation $A\otimes B$ means,%
\begin{equation}
\left( A\otimes B\right) \left( W_{f}^{j_{1}}W_{g}^{j_{2}}\right) =\left(
AW_{f}^{j_{1}}\right) \left( BW_{g}^{j_{2}}\right) .  \label{tp}
\end{equation}

Explicitly applying Equation~\eqref{exp} to the symbols of operators $\hat{f}
$ and $\hat{g}$\ and performing an integration and summation, one obtains the
following symbolic expression for the symbol of their product
\begin{equation}
W_{fg}^{j}\left( \Theta \right) \approx \mathcal{V}\left( W_{f}^{j+I\otimes
\mathbb{J}^{0}}\left( \Theta \right) W_{g}^{j-\mathbb{J}^{0}\otimes I}\left(
\Theta \right) \right) ,  \label{sps}
\end{equation}%
where the operator indices of each symbol are applied to the right or to the
left according to~\eqref{tp}. The above expression can be further simplified
in the limit $j\gg 1$ and considering $j$ as a continuous variable (see
Section~\ref{discretization}). However, the continuous limits for symbols $%
W_{f}^{j\ }\left( \Theta \right) $ are different for even and odd \ values
of the index $j$ due to the parity property~\eqref{we} and \eqref{wo}. It is
convenient to introduce the linear combinations
\begin{eqnarray}
W_{f}^{j+\ }\left( \Theta \right) &=&W_{f}^{j\ }\left( \Theta \right)
+W_{f}^{j+1\ }\left( \Theta \right) ,  \label{W+} \\
W_{f}^{j-\ }\left( \Theta \right) &=&\left( -1\right) ^{j}\left( W_{f}^{j\
}\left( \Theta \right) -W_{f}^{j+1\ }\left( \Theta \right) \right) ,
\label{W-}
\end{eqnarray}%
which are related through a phase shift,
\begin{equation}
W_{f}^{j-\ }(\phi ,\theta ,\psi )=W_{f}^{j+\ }(\phi ,\theta ,\psi +\pi ).
\label{wpm}
\end{equation}%
For instance, $W_{r_{k}}^{j+}(\Theta )=r_{k},$ for any (integer) value of
the index $j$. The symbols~\eqref{W+} and \eqref{W-} become smooth functions of
$j$ in the continuous limit, $W_{f}^{j\pm \ }\left( \Theta
\right)\rightarrow W_{f}^{\pm \ }\left( \Theta ,j\right) $.

Of particular interests are symbols $W^{j\pm \ }\left( \Theta \right) $ with
index $j$ distributed in a vicinity, $1\ll \delta \ll j_{0},$ of some $j_{0}$%
. Actually, smooth and localized functions of $\left( \Theta ,j\right) $,
with $\delta \sim j_{0}^{1/2}$ for $j_{0}\gg 1$ correspond to the so-called
semi-classical states. Physically, such states are spread among several $%
SO(3)$ invariant subspaces characterized by a large value of spin and
localized in angle variables.

The Schrodinger equation%
\[
i\partial _{t}\hat{\rho}=[\hat{H},\hat{\rho}],
\]%
$\hat{H}$ being the Hamiltonian of the system, is mapped into the evolution
equations for the Wigner functions
\begin{equation}
i\partial _{t}W_{\rho }^{j}=W_{H\rho }^{j}-W_{\rho H}^{j}.  \label{Weq}
\end{equation}

In the continuous limit and for initial semi-classical states, Equation~%
\eqref{Weq} is reduced in the leading order on $j_{0}$ to the
Liouville-type differential equations \cite{Tomatani(2015)} (see also
\mbox{Appendix~\ref{AppendixA}}) for $W_{\rho }^{\ \pm }\left( \Theta ,j\right) $,
\begin{equation}
\partial _{t}W_{\rho }^{\ \pm }\left( \Theta ,j\right) \approx 2\{W_{H}^{\
\pm }\left( \Theta ,j\right) ,W_{\rho }^{\ \pm }\left( \Theta ,j\right) \},
\label{ee}
\end{equation}%
where $\{.,.\}$ are the Poisson brackets in the Darboux coordinates $\left(
\left( j+1\right) \cos \theta ,\phi \right) $ and $\left( j,\psi \right) $,
and $W_{H}^{\pm }$ are the corresponding symbols of the Hamiltonian. The
above equation defines a classical evolution on the symplectic manifold
isomorphic to the cotangent bundle $T^{\ast }\mathcal{S}_{2}$, which
corresponds to the co-adjoint orbit of the $SE(3)$ group fixed by the values
of the Casimir operators $\mathbf{\hat{r}}^{2}=\hat{I}$ and $\mathbf{\hat{l}}%
\cdot \mathbf{\hat{r}}=0$. Thus, in the semi-classical limit, the Wigner
functions $W_{\rho }^{\pm }(\Theta ,j)$ can be considered as distributions
in a four-dimensional manifold $T^{\ast }\mathcal{S}_{2}$ and Equation~%
\eqref{ee} determines \textquotedblleft classical
trajectories\textquotedblright\ $(\Theta ^{cl}(t),j^{cl}(t))$ for variable
spin systems. A classical observable $f$ can be associated either with $%
W_{f}^{+}\left( \Theta ,j\right) $ or $W_{f}^{-}\left( \Theta ,j\right) $;
however, it is more convenient to choose $W_{f}^{+}\left( \Theta ,j\right) $
due to the relation~\eqref{wpm}. The explicit form of the Poisson brackets
on $T^{\ast }\mathcal{S}_{2}$ is given in Appendix~\ref{AppendixA}, Equation~%
\eqref{PB}. Strictly speaking, the real expansion parameter, used in
transition from \eqref{Weq} to \eqref{ee}, is $j(t)^{-1}$ (for $j$ initially
localized close to $j_{0}\gg 1$). Therefore, the Liouville Equation %
\eqref{ee} holds, while, on average over the distribution, $j(t)\sim
\left\langle \mathbf{\hat{l}}^{2}\right\rangle ^{1/2}\gg 1$. 

The evolution of average value of an operator $\hat{f}$ evaluated according
to the overlap relation~\eqref{over} is convenient to rewrite in terms of
symbols $W_{f}^{j\pm \ }\left( \Theta \right) $, as follows:

\begin{equation}
\left\langle \hat{f}\left( t\right) \right\rangle =\sum_{j=0,1,2,\dots}\frac{%
j+1}{16\pi ^{2}}\int d\Theta W_{f}^{j+\ }\left( \Theta |t\right) \left(
W_{\rho }^{j+}\left( \Theta \right) +\left( -1\right) ^{j}W_{\rho
}^{j-}\left( \Theta \right) \right) ,  \label{fte}
\end{equation}%
where $W_{f}^{j+\ }\left( \Theta |t\right) $ is the symbol of the Heisenberg
operator $\hat{f}\left( t\right) $. In the continuous limit, changing
summation on $j$ to integration, the contribution of the second term in the
above equation becomes negligible. Then, in the spirit of TWA, considering $%
(\Theta ,j)$ as dynamic variables, the average values of physical
observables can be estimated according to~\eqref{over}

\begin{equation}
\langle \hat{f}(t)\rangle \approx \int_{0}^{\infty }dj\frac{j+1}{16\pi ^{2}}%
\int d\Theta W_{f}^{+\ }(\Theta ^{cl}(t),j^{cl}(t))W_{\rho }^{+}\left(
\Theta ,j\right) ,  \label{fta}
\end{equation}%
where the integration is carried out on the initial conditions of the
classical trajectories $j^{cl}(t)=j^{cl}(\Theta ,j|t)$, $\Theta
^{cl}(t)=\Theta ^{cl}(\Theta ,j|t)$.

Actually, since it is expected that $W_{\rho }^{+}(\Theta ,j)$ is sharply
localized at $j\sim j_{0}\gg 1$ (commonly the width of the \textit{%
semiclassical} Wigner distribution is $\sim \sqrt{j_{0}}$), Equation~%
\eqref{fta} can be well approximated as%
\begin{equation}
\langle \hat{f}(t)\rangle \approx \int_{-\infty }^{\infty }dj\frac{j+j_{0}+1%
}{16\pi ^{2}}\int d\Theta W_{f}^{+}\left( \Theta ,j+j_{0}|t\right) W_{\rho
}^{+}(\Theta ,j+j_{0}),  \label{fta1}
\end{equation}%
where
\[
W_{f}^{+}\left( \Theta ,j+j_{0}|t\right) \equiv W_{f}^{+}\left( \Theta
^{cl}(\Theta ,j+j_{0}|t),j^{cl}(\Theta ,j+j_{0}|t)\right) ,
\]%
and $W_{\rho }^{+}(\Theta ,j+j_{0})$ is now centered at zero in the $j$
axis. In addition, the semiclassical distributions $W_{\rho }^{j\pm }\left(
\Theta \right) $ are approximately normalized
\[
\sum_{j=0,1,2,\ldots }^{\infty }\frac{j+1}{16\pi ^{2}}\int d\Theta W_{\rho
}^{j\pm }\left( \Theta \right) \approx 1.
\]

As well as in cases of lower dimensional manifolds, it cannot be expected
that propagation along distinguishable classical trajectories (there is no
trajectory crossing), originated at every phase-space point of the initial
distribution, are able to describe a long-time non-harmonic dynamics \cite%
{Steuernagel(2013), Oliva(2018), Oliva(2019)}. However, as it will be shown
below, there is a natural form to improve the time-validity of TWA for
initial semi-classical states of variable spin systems.

It is worth emphasizing that only $W_{\rho }^{\ \pm }\left( \Theta ,j\right) 
$ combinations satisfy the Liouville evolution Equation (\ref{ee}). Thus,
the symbols $W_{f}^{+\ }\left( \Theta ,j|t\right) $ should be associated
with the evolving classical observables according to (\ref{fte}) and (\ref%
{fta}).

\section{Asymptotic Quantization and Discretization Procedure}

\label{discretization}

We start noting that the integral~\eqref{fta} may describe only\ a
destructive dynamic interference corresponding to the initial stage of
non-harmonic quantum evolution (collapse time). Such a behavior is due to a
\textit{continuous} superposition of independent classically propagated
infinitesimally close fractions of the initial distribution. Several
discretization \mbox{procedures \cite{Littlejohn(2002), Light(2000)}} have
been proposed in order to overcome this problem and to extend the time
validity of TWA. Among them, one can mention a semi-classical discretization
based on propagating a single trajectory out of each Plank cell in a flat %
\mbox{phase-space \cite{Takahashi(1993)}}, application of the discrete
Wigner function\ method \cite{Wootters(1987)} to a collection of $1/2$-spin 
\mbox{systems \cite{Schachenmayer_Many(2015), Schachenmayer_Dynamics(2015),
Acevedo(2017), Pineiro(2017), Pucci(2016), Sundar(2019)}.}

The form of phase-space evaluation of average values in variable spin system~%
\eqref{fte} suggests an intuitive way for a partial discretization of the
initial distribution in the semi-classical limit. In our approximation, the
discrete index $j$, which originally appeared in the expansion~\eqref{fj}
for labeling the angular momentum sectors, is now considered as a classical
dynamic variable. Then, having applied the star-product~\eqref{sps} to the
symbols~\eqref{W+} and \eqref{W-}, we immediately arrive in the continuous
limit at the following asymptotic form of the star-product 
\begin{eqnarray}
W_{fg}^{\pm }\left( \Theta ,j\right)  &\approx &e^{-\varepsilon \mathbb{J}%
^{0}\otimes \mathbb{J}^{0}-\frac{\varepsilon }{2}\left( \mathbb{J}%
^{+}\otimes \mathbb{J}^{-}-\mathbb{J}^{-}\otimes \mathbb{J}^{+}\right)
}e^{\partial _{j}\otimes \mathbb{J}^{0}-\mathbb{J}^{0}\otimes \partial _{j}}%
\left[ W_{f}^{\pm }\left( \Theta ,j\right) W_{g}^{\pm }\left( \Theta
,j\right) \right]   \label{sa} \\
&\approx &W_{f}^{\pm }\left( \Theta ,j\right) \ast W_{g}^{\pm }\left( \Theta
,j\right) ,
\end{eqnarray}%
where $e^{\partial _{j}\otimes \mathbb{J}^{0}}f(j)=f(j+I\otimes \mathbb{J}%
^{0})$. It is worth noting that the star-product in the form \eqref{sa} is
applicable only to the classical observables, i.e. \eqref{W+} and \eqref{W-}
symbols.

Strictly speaking, Equation~\eqref{sa} should be applied to functions with
the index $j$ distributed in a broad vicinity, $\delta \gg 1,$ of some $%
j_{0}\gg 1$, so that $\varepsilon \approx (j_{0}+1)^{-1}$ and $\partial
_{j}\sim \delta ^{-1}$. However, a direct application of Equation~\eqref{sa}
to the generators of the $se(3)$ algebra \linebreak\mbox{($\mathbf{\hat{l}}$, $\mathbf{\hat{r%
}}$)} shows a good correspondence between the exact commutation relations and
their counterparts reconstructed through the asymptotic star-product:%
\begin{eqnarray*}
W_{[l_{k},l_{n}]}^{+}\left( \Theta ,j\right) &=&i\epsilon
_{knm}W_{l_{m}}^{+}\left( \Theta ,j\right) +O\left( \varepsilon \right) \\
W_{[l_{k},r_{n}]}^{+}\left( \Theta ,j\right) &{=}&i\epsilon
_{knm}W_{r_{m}}^{+}\left( \Theta ,j\right) +O(\varepsilon ),\quad
W_{[r_{k},r_{n}]}^{+}\left( \Theta ,j\right) =0,
\end{eqnarray*}%
where
\begin{equation}
W_{[f,g]}\left( \Theta ,j\right) =W_{f}\left( \Theta ,j\right) \ast
W_{g}\left( \Theta ,j\right) -W_{g}\left( \Theta ,j\right) \ast W_{f}\left(
\Theta ,j\right) .  \label{com}
\end{equation}

It is worth noting that asymptotically the operator corresponding to the
classical observable $(j+j_{0})/2$ is conjugated to the operator
corresponding to the phase $e^{i\psi }$,
\begin{equation}
W_{[(j+j_{0})/2,e^{i\psi }]}^{+}\left( \Theta ,j\right) \approx W_{e^{i\psi
}}^{+}\left( \Theta ,j\right) .  \label{jpsi}
\end{equation}%
In a sense, $(j+j_{0})/2$ and $e^{i\psi }$ can be seen as action-angle
variables. Here, we consider $\left( j+1\right) /2$ as a physical variable
representing the classical angular momentum, as $W_{l_{k}}^{+}(j,\Theta
)\approx \frac{j+1}{2}\,n_{k}$, according to~\eqref{Wl} and \eqref{l2}.

Making use of the asymptotic form of the star-product~\eqref{sa}, we can
discretize back the variable $j$ following the ideas of deformation
quantization. According to the general procedure~\cite{Bayen(1978)}, we solve
the eigenvalue equation%
\[
i\partial _{\tau }U=\left( j+j_{0}\right) \ast U\approx \left(
j+j_{0}-i\partial _{\psi }\right) U,
\]%
\[
U\left( \left. j+j_{0},\Theta \right\vert \tau =0\right) =1,
\]%
where the expression~\eqref{sa} was employed. A direct expansion of the
solution

\[
U\left( \left. j+j_{0},\Theta \right\vert \tau \right) =e^{-i\ \left(
j+j_{0}\right) \tau },\qquad
\]%
in the Fourier series yields

\[
e^{-i\left( j+j_{0}\right) \tau }=\sum_{L=-\infty }^{\infty }\Pi
_{L}(j)e^{m\pi i\left( j+j_{0}\right) }e^{-i\left( L+j_{0}\right) \tau
},\qquad \left( 2m-1\right) \pi <\tau <\left( 2m+1\right) \pi ,
\]%
where
\begin{equation}
\Pi _{L}(j)=\frac{\sin \pi \left( j-L\right) }{\pi \left( j-L\right) }.
\label{Pi}
\end{equation}

Since the Wigner distributions $W_{\rho }^{\pm }\left( \Theta
,j+j_{0}\right) $ have compact supports, with width $\sim \sqrt{j_{0}}\gg 1$%
, we can make use of the Whittaker--Shannon--Kotelnikov sampling %
\mbox{theorem \cite{Whittaker(1915), Shannon(1949), Kotelnikov(1933)}},
approximating%
\begin{eqnarray}
&&\frac{j+j_{0}+1}{16\pi ^{2}}W_{f}^{+}\left( \Theta ,j+j_{0}|t\right)
W_{\rho }^{\pm }\left( \Theta ,j+j_{0}\right)  \label{sampl} \\
&\approx &\sum_{L=-\infty }^{\infty }\frac{L+j_{0}+1}{16\pi ^{2}}%
W_{f}^{+}\left( \Theta ,L+j_{0}|t\right) W_{\rho }^{\pm }\left( \Theta
,L+j_{0}\right) \ \Pi _{L}(j).  \nonumber
\end{eqnarray}%
It is worth noting that, in case of Gaussian function of width $\sim \sqrt{%
j_{0}}$, the error of discrete sampling with~\eqref{Pi} is of order $\sim $%
\textrm{erfc}$\left( \pi \sqrt{j_{0}}\right) $~\cite{Rybicki(1989)}.

\begin{figure}[H]%
\widefigure
\includegraphics[width=15 cm]{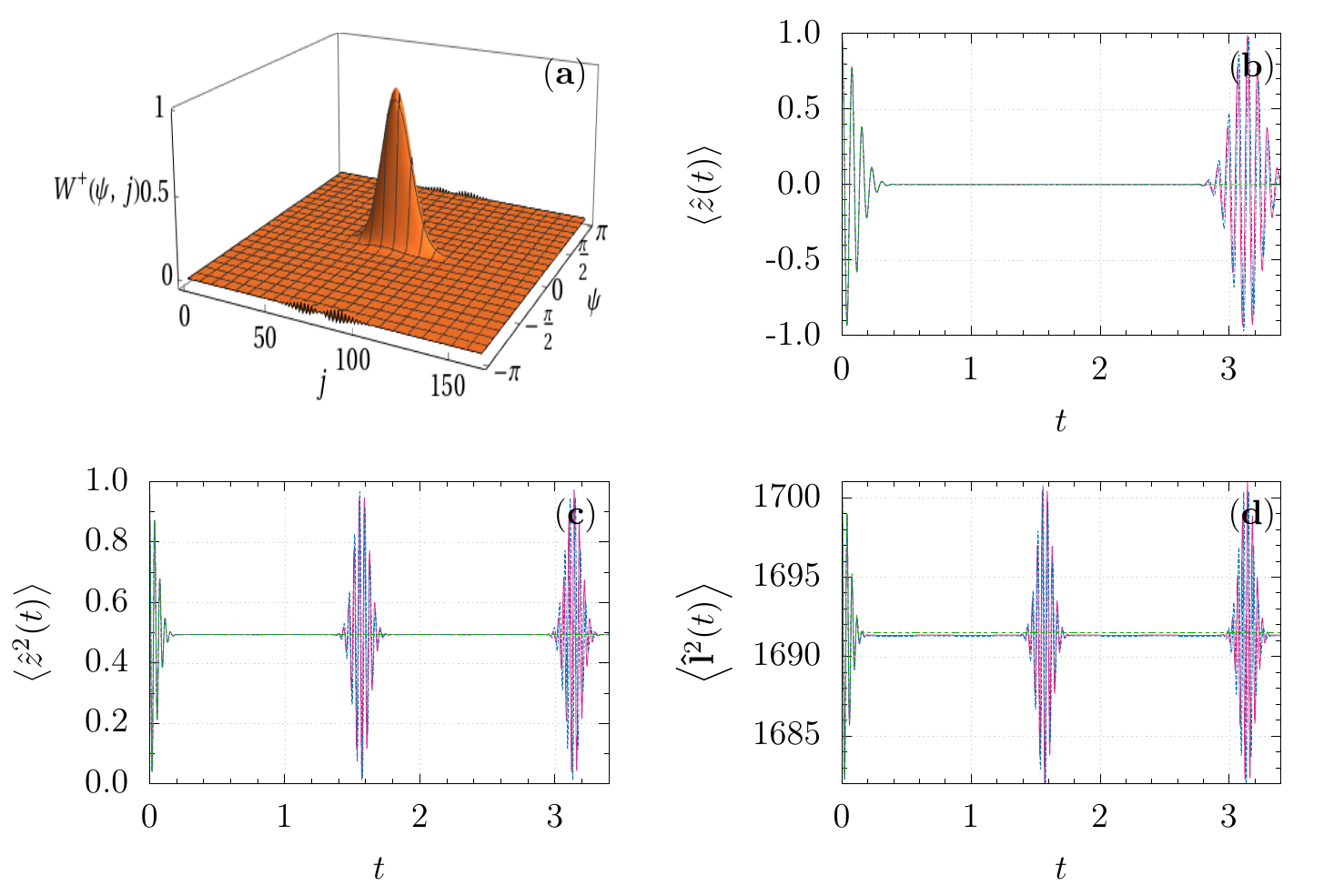}
\caption{(\textbf{a}) The marginal distribution $W_{\protect\rho }^{+}\left(
\protect\psi ,j\right) $ corresponding to the state \protect\eqref{psiR};
(\textbf{b}-\textbf{d}) evolution of $\langle \hat{z}(t)\rangle $, $\langle \hat{z}%
^{2}(t)\rangle $,   $\langle \mathbf{\hat{l}}^{2}(t)\rangle $  generated by
the Hamiltonian \protect\eqref{Hr1} with $g=20$, for the initial state~%
\protect\eqref{psiR} with $r^{2}=81$, $\protect\varphi _{0}=\protect\psi %
_{0}=0$, $\protect\vartheta _{0}=\protect\pi /2$: exact evolution (solid
magenta line), continuous  $T^{\ast }\mathcal{S}_{2}$ TWA~\protect\eqref{fta}
(dashed green line), discrete TWA \protect\eqref{DTWA} (dashed blue
line). }
\label{fig:rotor}
\end{figure}

However, a direct discretization~\eqref{sampl} of the semi-classical
expression~\eqref{fta1} is not sufficient for an efficient simulation of the
quantum dynamics through phase-space trajectories since the overlap between
the classically evolved observable $W_{f}^{j+\ }\left( \Theta |t\right) $
and the branch $W_{\rho }^{j-}\left( \Theta \right) $ of distribution would
be missed. It is worth recalling that $W_{\rho }^{+}\left( \Theta ,j\right) $
and $W_{\rho }^{-}\left( \Theta ,j\right) $ have maxima at different points
of the phase space (shifted in $\pi $ on $\psi $). In order to correct this
problem, we rewrite Equation~\eqref{fte}
\begin{eqnarray}
\left\langle \hat{f}\left( t\right) \right\rangle  &=&\Sigma _{+}(t)+%
\mathcal{P}\Sigma _{-}(t)  \label{fte1} \\
\Sigma _{\pm }(t) &=&\sum_{j=0,1,2,\dots}\frac{j+1}{16\pi ^{2}}\int d\Theta
W_{f}^{j+\ }\left( \Theta |t\right) W_{\rho }^{j\pm }\left( \Theta \right) ,
\label{S}
\end{eqnarray}
where $\mathcal{P}$ is the parity operator defined according to
\[
\mathcal{P}\sum_{j=0,1,2,..}a_{j}=\sum_{j=0,1,2,..}(-1)^{j}a_{j}.
\]%
Then, considering the semi-classical evolution of $W_{f}^{+\ }\left( \Theta
,j\right) $ in the continuous limit and applying a subsequent discretization
procedure~\eqref{sampl}--\eqref{S}, we arrive at the following discretized
version of TWA for variable spin systems,
\end{paracol}
\begin{equation}
\langle \hat{f}(t)\rangle \approx \sum_{L=-\infty }^{\infty }\frac{L+j_{0}+1%
}{16\pi ^{2}}\int d\Theta W_{f}^{+}\left( \Theta ,L+j_{0}|t\right) \left(
W_{\rho }^{+}\left( \Theta ,L+j_{0}\right) +(-1)^{L}W_{\rho }^{-}\left(
\Theta ,L+j_{0}\right) \right) .  \label{DTWA}
\end{equation}
\begin{paracol}{2}
\switchcolumn

The above equation is just a convolution of the evolving classical
observable with a linear combination of the initial distribution $W_{\rho
}^{+}$ at $(\phi ,\theta ,\psi )$ and $(\phi ,\theta ,\psi +\pi )$,
evaluated at the equiseparated points along the ``action'' variable $j$. A
naive direct discretization of the continuous approximation (\ref{fta})
leads to an incomplete description of the quantum dynamics in the
semiclassical limit.

\section{Examples}

\label{examples}

\subsection{Rigid Rotor in an External Field}

Let us consider the following Hamiltonian governing the evolution of quantum
rigid rotor in an external field along the $z$-axis, 
\begin{equation}
\hat{H}=\mathbf{\hat{l}}^{2}-g\hat{z}^{2},  \label{Hr1}
\end{equation}%
where $\hat{z}$ is the $z$-component of the orientation operator (\ref{nz0}%
), and $\mathbf{\hat{l}}^{2}$ is the square angular momentum operator. The
Hamiltonian (\ref{Hr1}) possesses the $SO(3)$ symmetry but cannot be reduced
to a finite number of spin systems.

The symbol of the Hamiltonian in the continuous limit at the principal order
on $j$ has the form%
\begin{equation}
W_{H}^{+}(\Theta ,j)\approx \frac{j}{2}\left( \frac{j}{2}+1\right) -g\sin
^{2}{\theta }\cos ^{2}{\psi }
\end{equation}%
and leads to the following equations of motion:
\begin{eqnarray}
\partial _{t}\phi  &=&\frac{4g}{j+1}\cos {\theta }\cos ^{2}{\psi } \\
\partial _{t}\theta  &=&-\frac{g}{j+1}\sin {2\theta }\sin {2\psi }  \nonumber
\\
\partial _{t}\psi  &=&j+1-\frac{4g}{j+1}\cos ^{2}{\theta }\cos ^{2}{\psi }
\nonumber \\
\partial _{t}j &=&-2g\sin ^{2}{\theta }\sin {2\psi }.  \nonumber
\end{eqnarray}%
As an initial state, we consider a weighted superposition of $l$-spin
coherent states $\left\vert l;\vartheta _{0},\varphi _{0}\right\rangle $
\begin{equation}
\left\vert \Psi (0)\right\rangle =\frac{1}{\sqrt{\cosh {r^{2}}}}%
\sum_{l=0}^{\infty }e^{-il\psi _{0}}\frac{r^{2l}}{\sqrt{(2l)!}}\left\vert
l;\vartheta _{0},\varphi _{0}\right\rangle ,  \label{psiR}
\end{equation}%
where $r^{2}\gg 1$. The Wigner distribution $W_{\rho }^{+}\left( \Theta
,j\right) $ corresponding to the state~\eqref{psiR} can be approximated as%
\begin{eqnarray*}
W_{\rho }^{+}\left( \Theta ,j\right)  &\approx &\frac{r^{2j}}{\Gamma \left(
j+2\right) \cosh r^{2}}\frac{1}{\sin \omega /2}\partial _{\omega }\left(
\frac{\sin ^{2}\left( \left( j+1\right) \omega /2\right) }{\sin \omega /2}%
\right) , \\
\cos \frac{\omega }{2} &=&\frac{1}{\sqrt{2}}\left( \cos \frac{\theta }{2}%
\cos \frac{\phi +\psi }{2}+\sin \frac{\theta }{2}\cos \frac{\phi -\psi }{2}%
\right) ,
\end{eqnarray*}%
and is localized in $\vartheta \sim \vartheta _{0}=\pi/2$, $\varphi \sim
\varphi _{0}=0$ by construction, with absolute fluctuations $\delta
\vartheta \sim \delta \varphi \sim r^{-1}$. In addition, $W_{\rho
}^{+}\left( \Theta ,j\right) $ is localized at $j\sim r^{2}$ with the
fluctuation $\delta j\sim r$. Thus, the phase $\psi $ is also localized with 
$\delta \psi \sim r^{-1}$, as it is an observable conjugated to $j$,
according to~\eqref{jpsi}, see Figure~\ref{fig:rotor}a where the marginal
distribution%
\begin{equation}
W_{\rho }^{+}\left( \psi ,j\right) =\frac{j+1}{8\pi }\int_{0}^{2\pi }d\phi
\int_{0}^{\pi }\sin \theta d\theta W_{\rho }^{+}\left( \Theta ,j\right) 
\label{Wjp}
\end{equation}%
is plotted. The same reasoning is applicable to $W_{\rho }^{-}\left( \Theta
,j\right) $ due to the relation~\eqref{wpm}. This means that the state~%
\eqref{psiR} is a semi-classical state on the manifold $T^{\ast }\mathcal{S}%
_{2}$.

We have numerically tested the approximation~\eqref{DTWA} by computing the
averages of $\hat{z}(t)$, $\hat{z}^{2}(t)$, and $\mathbf{\hat{l}}^{2}(t)$.
For numerical simulations, we have used an adaptive sampling technique that
allows us to sample the angular variables in the regions where the initial
Wigner function is located, for every fixed value of $j$. The sampling
method is based on an algorithm \cite{Genz(1980)} for numerical integration
of multivariate functions~ and implemented by a modification of the routine 
\textit{cubature}~\cite{Johnson(2017)}. For $r^{2}=81$, $g=20$, we take on
average $7892$ points inside $S_{3}$ sphere for each value of integer $j\in %
\left[ 40,160\right] $. The errors obtained in the estimation of the
expectation values are lower than $0.0045\%$ at $t=0$. The differential
equations were solved by a variable step-size Runge--Kutta method of order $%
9(8)$~\cite{Verner(2010)}. The size of the step was adapted to keep the
relative errors estimated by the method lower \mbox{than $10^{-6}$.}

In Figure~\ref{fig:rotor}b--d, we plot the corresponding averages in
comparison with the exact calculations and the continuous TWA approximation~%
\eqref{fta1}. One can appreciate that Equation~\eqref{fta1} coincides with
the exact calculations only within the initial collapse, i.e., for times $%
gt\ll 1$. Conversely, Equation~\eqref{DTWA} describes very well the
evolution of the observables inclusively for much longer intervals that
include several revivals, $gt\sim \pi $. For even longer times, $gt\sim
r^{2} $, our approximation starts to deviate from the exact solution,
failing to capture the oscillation dephasing and deformations of envelopes
of the revivals (although the condition $j(t) \gg 1$ still holds).

It is worth noting that, in contrast to planar pendulum models,
described by periodic Hamiltonians of the form%
\[
\hat{H}=\hat{p}^{2}+V(\hat{x}),\quad V(x+a)=V(x),\quad a=const,
\]%
where $\hat{p}$ and $\hat{x}$ are the standard momentum and position
operators, a phase-space description of the rigid rotor in an external field
is not trivial. For instance, the standard phase-space analysis of the
Hamiltonian (\ref{Hr1}) in $C^{2}\otimes C^{2}$, and application of the
corresponding TWA \cite{Polkovnikov(2010)}, faces considerable technical
difficulties. In particular, the Schwinger (two-mode) representation of $%
\hat{z}$ operator,%
\[
\hat{z}=\frac{1}{\sqrt{\hat{a}\hat{a}^{\dagger }+\hat{b}^{\dagger }\hat{b}}}%
\left( \hat{a}^{\dagger 2}+\hat{b}^{\dagger }\right) \frac{1}{\sqrt{\hat{a}%
\hat{a}^{\dagger }+\hat{b}^{\dagger }\hat{b}}},
\]%
is quite inconvenient for the $H(1)\times H(1)$
phase-space mapping \cite%
{Glauber(1963),Sudarshan(1963),Cahill(1969)1857,Cahill(1969)1882}, and hides
the intrinsic $SO(3)$ symmetry of the direction operator (%
\ref{nz0}) and (\ref{x base}).

\end{paracol}
\nointerlineskip
\begin{paracol}{2}
\switchcolumn
\vspace{-9pt}
\subsection{Spin--Spin Interaction}

As another non-trivial example, we consider the following Hamiltonian:
\begin{equation}
\hat{H}=\mathbf{\hat{l}}_{1}\cdot \mathbf{\hat{l}}_{2}+\lambda \hat{l}_{z1},
\label{SS}
\end{equation}%
describing an integer spin--spin interaction in the presence of an external
non-uniform \mbox{magnetic field.}

Within the framework of our approach, the symbol of the Hamiltonian in the
continuous limit has the form%
\[
W_{H}^{+}(\Theta ,j)=C(j)+\lambda A(j)\sin \theta \cos \psi +\lambda
B(j)\cos \theta ,
\]%
where $A(j)$, $B(j)$ and $C(j)$ are functions of $j$ given in Appendix~\ref%
{AppendixB}.

Taking into account the classical equations of motion on $T^{\ast }\mathcal{S%
}_{2}$ (see Appendix~\ref{AppendixB}), we compute the evolution of the first
spin magnetization $\left\langle \hat{l}_{x1}(t)\right\rangle $ according to
the general procedure~\eqref{DTWA}, where the symbol of $\hat{l}_{x1}$ is%
\[
W_{l_{x1}}^{+}\left( \Theta ,j\right) =A(j)\left( \sin \phi \sin \psi -\cos
\phi \cos \theta \cos \psi \right) +B(j)\cos \phi \sin \theta . 
\]%
As the initial state, we consider the product of spin coherent states in $x$%
- and $y$-directions,%
\begin{equation}
|\Psi (0)\rangle =\left\vert L_{1};\varphi _{1}=0,\vartheta _{1}=\pi
/2\right\rangle \otimes \left\vert L_{2};\varphi _{2}=\pi /2,\vartheta
_{2}=\pi /2\right\rangle .  \label{psiSS}
\end{equation}%
The general expression for the Wigner function of the state~\eqref{psiSS} is
quite cumbersome, but its localization property at $j\sim 2L_{1}$ for $%
L_{1}\gg L_{2}$ follows from the marginal distribution 
\[
\int_{0}^{2\pi }\frac{d\psi }{2\pi }W_{\rho }^{+}\left( \Theta ,j\right)
\sim e^{-\left( j/2-L_{1}\right) ^{2}/L_{2}}W_{\rho }\left( \phi ,\theta
;j\right) , 
\]%
where $W_{\rho }\left( \phi ,\theta ;j\right) =\langle L=j/2;\varphi
=0,\vartheta =\pi /2|\hat{w}_{L=j/2}(\phi ,\theta )\left\vert L=j/2;\varphi
=0,\vartheta =\pi /2\right\rangle $ is the standard $SO(3)$ Wigner function~%
\eqref{wsu2}. The localization on the angle $\psi $ follows from its
complementarity to the variable $j$, in the same way as in the rotor case.
The marginal distribution~\eqref{Wjp} corresponding to the state~%
\eqref{psiSS} is plotted in Figure~\ref{fig:spin}a.\vspace{-9pt}

\begin{figure}[H]
\includegraphics[width=10 cm]{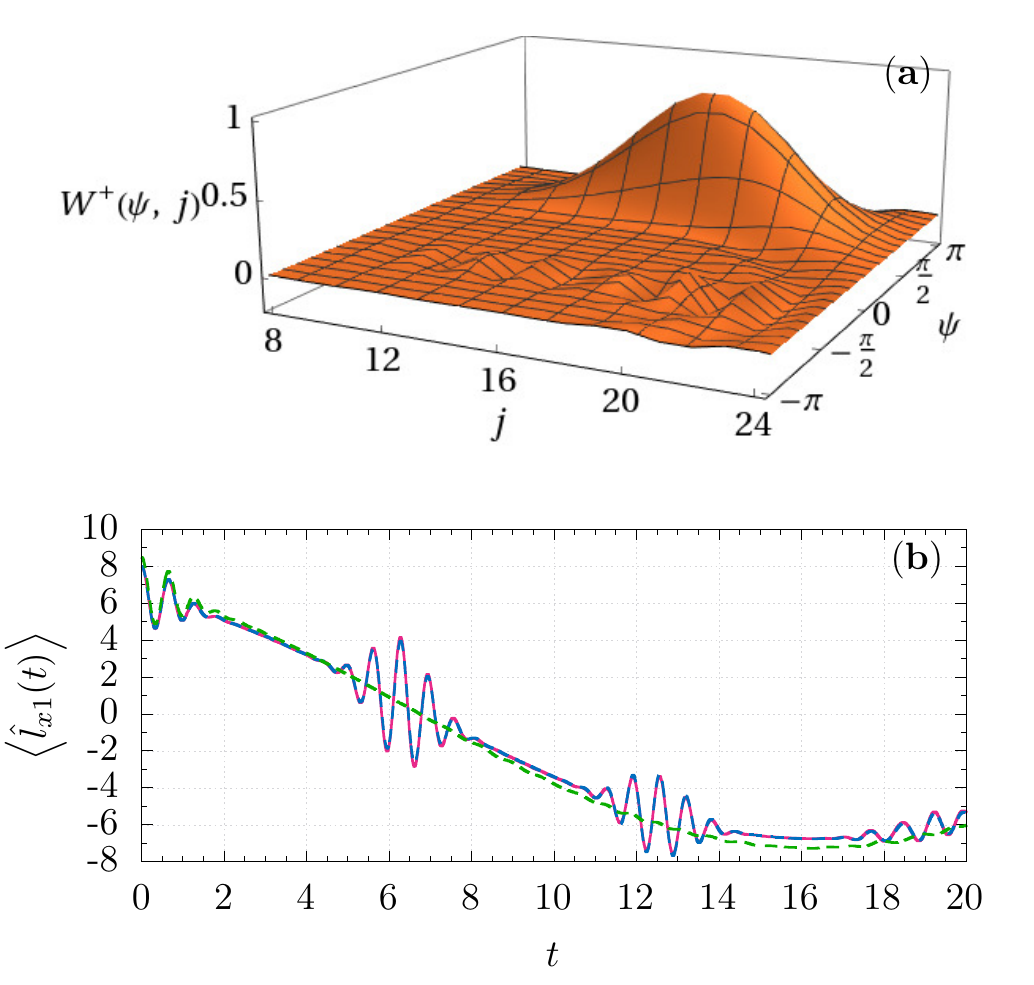}
\caption{(\textbf{a}) The marginal distribution $W_{\protect\rho }^{+}\left(
\protect\psi ,j\right) $ corresponding to the state~\eqref{psiSS}; (\textbf{%
b})~evolution of $\left\langle \hat{l}_{x1}(t)\right\rangle $ generated by
the Hamiltonian~\eqref{SS} with $\protect\lambda $ = 0.2, the initial state is
$\left\vert L_{1}=8;\protect\varphi _{1}=0,\protect\vartheta _{1}=\protect%
\pi /2\right\rangle \left\vert L_{2}=4;\protect\varphi _{2}=\protect\pi /2,%
\protect\vartheta _{2}=\protect\pi /2\right\rangle $: exact evolution (solid
magenta line), $SO(3)\times SO(3)$ TWA~\eqref{ftww} (dashed green line),
discrete TWA~\eqref{DTWA} (dashed blue line).}
\label{fig:spin}
\end{figure}

It is worth noting that this system has $SO(3)\times SO(3)$ symmetry (for
integer spins), so that the operators acting in the Hilbert space of
two-spin system can be mapped into distributions in $\mathcal{S}_{2}\times 
\mathcal{S}_{2}$ using the following the mapping kernel, 
\begin{equation}
\hat{w}(\Omega _{1},\Omega _{2})=\hat{w}_{1}(\Omega _{1})\otimes \hat{w}%
_{2}(\Omega _{2}),  \label{ww}
\end{equation}%
where $\hat{w}(\Omega )$, $\Omega =\left( \phi ,\theta \right) $ is defined
in~\eqref{wsu2}. In the limit of large spins, the Hamiltonian dynamics can
be treated semi-classically \cite{Amiet(1991), Klimov(2002), Klimov(2005),
Kalmykov(2016), Klimov(2017)}. The equation of motion for the Wigner
function of the whole system $\tilde{W}_{\rho }(\Omega _{1},\Omega _{2})$
takes the form:%
\begin{eqnarray*}
\partial _{t}\tilde{W}_{\rho } &\approx &\frac{1}{L_{1}+1/2}\{\tilde{W}_{H},%
\tilde{W}_{\rho }\}_{1}+\frac{1}{L_{2}+1/2}\{\tilde{W}_{H},\tilde{W}_{\rho
}\}_{2}, \\
&& \\
\tilde{W}_{H}(\Omega _{1},\Omega _{2}) &=&\sqrt{L_{1}(L_{1}+1)L_{2}(L_{2}+1)}%
\mathbf{n}_{1}\cdot \mathbf{n}_{2}+\lambda \sqrt{L_{1}(L_{1}+1)}n_{z1},
\end{eqnarray*}%
where the canonical variables defining the Poisson brackets $\{..,..\}_{1,2}$
are $\left( \cos \theta _{1,2},\phi _{1,2}\right) $, and the average values
are computed according to the standard TWA,%
\begin{equation}
\langle \hat{f}(t)\rangle \approx \frac{(2L_{1}+1)(2L_{2}+1)}{16\pi ^{2}}%
\int d\Omega _{1}d\Omega _{2}\tilde{W}_{f}((\Omega _{1}^{cl}(t),\Omega
_{2}^{cl}(t))\tilde{W}_{\rho }(\Omega _{1},\Omega _{2}),  \label{ftww}
\end{equation}%
$\Omega _{1}^{cl}(t)$ and $\Omega _{2}^{cl}(t)$ being classical trajectories
on $\mathcal{S}_{2}\times \mathcal{S}_{2}$.

For numerical simulations, we used the same adaptive method as in the case
of the rotor. For the spin--spin system, $L_{1}=8$, $L_{2}=4$, $\lambda =0.2$%
, $j\in \left[ 8,24\right] ,$ the average number of samples is 26,885. The
errors in the estimation of the expectation values are lower than $0.0008\%$
at $t=0$.

Comparing the results obtained from~\eqref{DTWA} and~\eqref{ftww}, one can
observe that the proposed discretization leads to a good coincidence with
the exact results significantly beyond the validity of the standard TWA in
the framework of Stratonovich--Weyl correspondence. Actually, our
approximation describes well the effect of partial revivals produced by the
nonlinear term $\hat{l}_{1}\cdot \hat{l}_{2}$ at the scale $t\sim 2\pi $,
but falling at $t\sim 2L_{1}$ (independently on the external field coupling
constant $\lambda $). The standard TWA breaks down already after the first
collapse at $t\lesssim 1$.

\section{Conclusions}

\label{conclusions}

The semi-classical map~\eqref{fW}--\eqref{j_sym} of density matrices of a
variable spin system into distributions on four-dimensional symplectic
manifold allows for approximating the evolution of such quantum systems in
terms of effective classical dynamics on $T^{\ast }\mathcal{S}_{2}$. The
advantage of the map~\eqref{j_sym} with respect to the standard $SU(2)/SO(3)$
case \cite{Stratonovich(1957)} consists of the possibility of a faithful
representation of operators whose action is not restricted to a single $%
SU(2) $ invariant subspace. In addition, only four Hamilton equations are
sufficient to determine the evolution of classical observables for any value
of the total angular momentum. However, the simplest Truncated Wigner
Approximation suffers from the same intrinsic defects as in the
Heisenberg--Weyl and $SU(2)$ symmetries: it describes well only the
short-time evolution of the typical observables of the system (which may
include, e.g., generators of $SE(3)$ group). In order to extend the validity
of TWA, while still keeping the idea of classical propagation, we propose to
``quantize'' back one of the classical dynamic variables, $j$, which can be
considered to some extent as an ``action'' (actually representing possible
values of the classical spin size). We perform such a ``quantization'' by
using the asymptotic form of the star-product within the framework of
deformation quantization, leading to a natural discretization of the
variable $j$, and, thus, all of the distributions appearing in the theory,
corresponding both to states and observables. A certain subtlety of the
proposed method consists of taking into account the parity problem
originated from the decomposition of the mapping kernel~\eqref{kernel} in
the basis of the tensor operators~\eqref{ITO}. In addition, it results that
the obtained discretization of initial distributions with compact support,
describing the so-called semi-classical states, is in direct accordance with
the famous sampling theorem. This allows the form of calculation of average
values to be immediately discretized, basically starting the classical
trajectories only at certain points of the initial distribution. The result
of such an approach is surprisingly good, as shown in Figures~\ref{fig:rotor}
and~\ref{fig:spin}. It is worth noting that the discrete sampling procedure
in general is not obvious at all. For instance, applying more sophisticated
discretization methods, like, e.g., the adaptive discretization, one obtains
much worse results than by following the simple recipe~\eqref{DTWA}.
Actually, Equation~\eqref{DTWA} describes very well all interference
effects, such as, e.g., revivals of quantum oscillations that appear due to
superpositions of subspaces with different values of the index $j$,
appearing in the exact calculations~\eqref{over}.

The range of applicability of the discretized TWA is considerably longer
than the standard semiclassical time $\tau _{sem}$, $gt\lesssim j_{0}^{\beta
}\tau _{sem}$, $\beta >0$, where $j_{0}\gg 1$ is the average initial total
angular moment. For second degree Hamiltonians, similar to (\ref{Hr1}) and (%
\ref{SS}), the leading corrections to the Liouville Equation (\ref{ee}) are
of the order $j_{0}^{-1}$, while the principal term (the Poisson bracket on $%
T^{\ast }\mathcal{S}_{2}$) is $\sim j_{0}$. Actually, the Moyal Equation (%
\ref{Weq}) has in this case the following structure%
\begin{equation}
\partial _{t}W_{\rho }^{\ }=\left( j_{0}\mathcal{L}_{0}+\mathbf{\ }j_{0}^{-1}%
\mathcal{L}_{2}\right) W_{\rho }^{\ }+O(j_{0}^{-2}),  \label{mea}
\end{equation}%
where $\mathcal{L}_{0}$ is a first order differential operator and $\mathcal{%
L}_{2}$ contains higher-degree derivatives. Dropping the correction terms $%
j_{0}^{-1}\mathcal{L}_{2}$ in the continuous TWA leads to neglecting all of
the commutators of order $j_{0}^{0}=1$ that appear in the exponent of the
formal propagator corresponding to Equation (\ref{mea}). As a result, the
description of quantum dynamics occurring in the time-scale $gt\gtrsim 1$ is
inaccessible in the semiclassical treatment (\ref{fta}). In practice, the
standard TWA \cite{Polkovnikov(2010)} breaks down even for shorter time
intervals and is unable to describe any genuine quantum effect such as, for
example, quantum revivals. It seems that the discretization \eqref{DTWA}
allows physical processes caused by the interferences between different $j$%
-sectors (\ref{fj}) to be emulated until $gt\lesssim j_{0}$. This is really
not surprising, since similar effects take place in almost all nonlinear
quantum systems with discrete spectra as a result of a specific composition
of some \textit{discrete} constituents \cite{Robinett(2004)}. The difficulty
consists, as we have mentioned above, in finding an appropriate
discretization of the classical phase-space. In Figure~\ref{fig:LongTime}%
a,b, the long time evolution of the rotor (\ref{Hr1}) and spin observables (%
\ref{SS}) are shown. One can clearly observe the region of time validity of
our approximation.

\end{paracol}
\nointerlineskip
\begin{figure}[H]%
\widefigure
\includegraphics[width=15 cm]{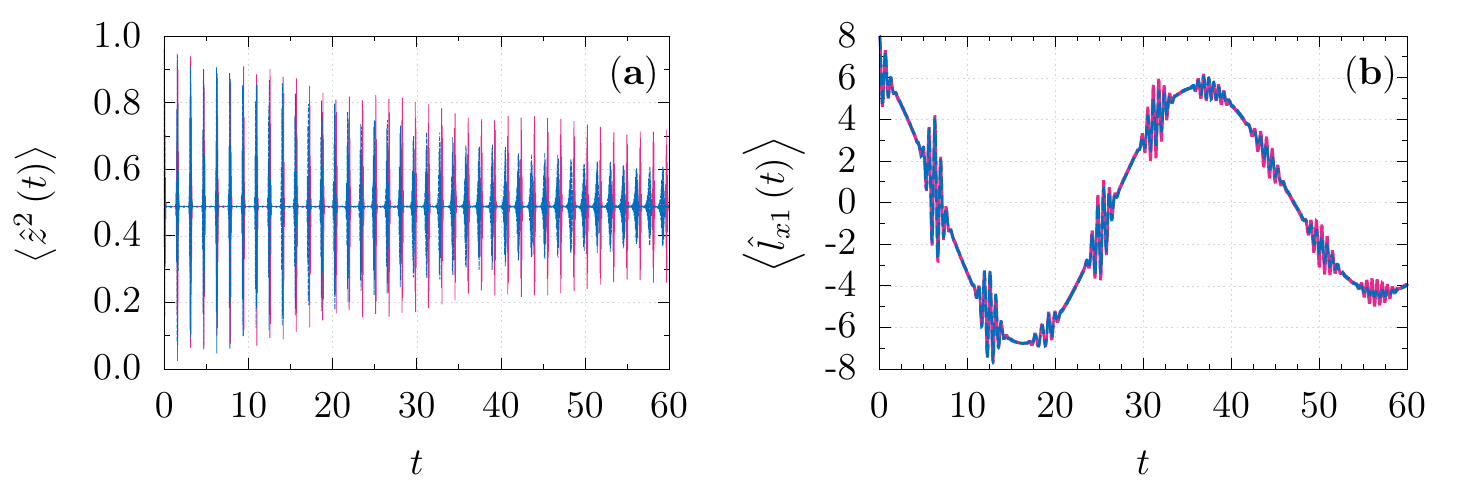}
\caption{(\textbf{a}) Long-time evolution of $\left\langle \hat{z}^{2}(t)\right\rangle $
generated by the Hamiltonian (\ref{Hr1}) with $g=20$, for the initial \mbox{state (\ref{psiR})} with $r^{2}=49$, $\phi =0,\psi =0,\theta =\pi /2$;
(\textbf{b}) evolution of $\left\langle \hat{l}_{x1}(t)\right\rangle $ generated by the
Hamiltonian~\eqref{SS} with $\lambda $ = 0.2, the initial state is $\left\vert
L_{1}=8;\varphi _{1}=0,\vartheta _{1}=\pi /2\right\rangle \left\vert
L_{2}=4;\varphi _{2}=\pi /2,\vartheta _{2}=\pi /2\right\rangle $: exact
evolution (solid magenta line), discrete \mbox{TWA~\eqref{DTWA} (dashed blue line).}}
\label{fig:LongTime}
\end{figure}
\begin{paracol}{2}
\switchcolumn

The present approach is also extendable to half-integer spins, although the
calculations become more involved. In such a case, four linear combinations
of the Weyl symbols $W_{f}^{j\ }\left( \Theta \right),$ similar to (\ref{W+}%
) and (\ref{W-}), should be introduced in order to follow the same procedure
as in Sections~\ref{semiclassical limit} and \ref{discretization}. This
problem will be considered elsewhere in application to dissipative and
pumped down conversion processes.

Recently, a generalized TWA was proposed in \cite{Zhu(2019)} for the
description of the dynamics of many coupled spins. In this approach, every
spin is considered as a discrete variable, i.e., there is no relation to any
classical phase-space. Thus, a set of $2\left( \left( 2L+1\right)
^{2}-1\right) $ coupled (first-order) differential equations should be
solved even for two interacting spins $L$. In addition, the rigid rotor
evolution in external fields cannot be treated applying the technique \cite%
{Zhu(2019)}.

Unfortunately, the map (\ref{fW})--(\ref{kernel}) cannot be used as a
faithful (one-to-one map) classical representation of the $N$-spin, $N\geq 3$
system. However, some global properties of multi-spin systems can be
analyzed with our method by using the Schur--Weyl duality \cite%
{Goodman(1998)} and averaging over invariant subspaces of the same
dimensions.

Finally, we note that the developed approach can, in principle, be applied
to any of the $s$-parametrized maps \cite{Tomatani(2015)}. However, the
Moyal equation for non self-dual distributions $W_{\rho }^{(s)\ }$ contains
terms of order one (on the semiclassical parameter), i.e., it has the form%
\[
\partial _{t}W_{\rho }^{(s)\ }=\left( j_{0}\mathcal{L}_{0}+s\mathcal{L}_{1}+%
\mathbf{\ }j_{0}^{-1}\mathcal{L}_{2}\right) W_{\rho }^{(s)\ }+O(j_{0}^{-2}), 
\]%
where $\mathcal{L}_{1}$ is a differential operator already containing higher
than first degree derivatives. Thus, one may expect that the TWA for $%
W_{\rho }^{(s\neq 0)\ }$, which considers the action only of $\mathcal{L}%
_{0} $, is less precise than for $W_{\rho }^{(s=0)\ }$.
\vspace{6pt}



\authorcontributions{Conceptualization, A.B.K. and J.L.R.; methodology,
  A.B.K. and J.L.R.; software, G.E.M.-H. and J.C.C.; validation, A.B.K., J.L.R.,
  G.E.M.-H., and J.C.C.; formal analysis, A.B.K., J.L.R., G.E.M.-H., and J.C.C.;
  investigation, A.B.K., J.L.R., G.E.M.-H., and J.C.C.;
  data curation, G.E.M.-H. and J.C.C.; writing---original draft preparation,
  A.B.K. and J.L.R.; writing---review and editing, A.B.K., J.L.R., G.E.M.-H., and
  J.C.C.; visualization, G.E.M.-H. and J.C.C.; supervision, A.B.K. All
  authors have read and agreed to the published version of the
  manuscript.}

\funding{This work is partially supported by the Grant 254127 of CONACyT
(Mexico).}

\institutionalreview{Not applicable.}

\informedconsent{Not applicable.}

\dataavailability{Not applicable.}

\acknowledgments{The authors are grateful for the computational resources and
technical support offered by the Data Analysis and Supercomputing Center
(CADS, for its acronym in Spanish) through the ``Leo Atrox'' supercomputer of
the University of Guadalajara, Mexico.}

\conflictsofinterest{The authors declare no conflict of interest.}



%
%
\appendixtitles{no}
\appendixstart
\appendix

\section{}\label{AppendixA}

Since the continuous limits are different for symbols $W_{f}^{j\ }\left(
\Theta \right) $ with even and odd values of the index $j$, Equation~%
\eqref{sps} should be expanded separately for $W_{fg}^{j\ (e)}\left( \Theta
\right) $ and $W_{f}^{j(o)\ }\left( \Theta \right) .$ For even values of the
index $j$, Equation~\eqref{sps} takes the form%
\begin{equation}
W_{fg}^{j\ (e)}\left( \Theta \right) \approx \mathcal{V}\left(
W_{f}^{j+I\otimes \mathbb{J}^{0}\ (e)}\left( \Theta \right) W_{g}^{j-\mathbb{%
J}^{0}\otimes I\ (e)}\left( \Theta \right) +W_{f}^{j+I\otimes \mathbb{J}%
^{0}\ (o)}\left( \Theta \right) W_{g}^{j-\mathbb{J}^{0}\otimes I\ (o)}\left(
\Theta \right) \right) ,  \label{pp}
\end{equation}%
while, for odd $j$, it becomes
\begin{adjustwidth}{-4.6cm}{0cm} 
\begin{equation}
W_{fg}^{j+1\ (o)}\left( \Theta \right) \approx \mathcal{V}\left(
W_{f}^{j+1+I\otimes \mathbb{J}^{0}\ (e)}\left( \Theta \right) W_{g}^{j+1-%
\mathbb{J}^{0}\otimes I\ (o)}\left( \Theta \right) +W_{f}^{j+1+I\otimes
\mathbb{J}^{0}\ (o)}\left( \Theta \right) W_{g}^{j+1-\mathbb{J}^{0}\otimes
I\ (e)}\left( \Theta \right) \right) ,  \label{pi}
\end{equation}
\end{adjustwidth}
where $\mathcal{V}$ is defined in~\eqref{V}.

In the semi-classical limit, when $j\gg 1$ is considered continuous, the
direct expansion of the above equations gives
\begin{eqnarray}
W_{f}^{j+I\otimes \mathbb{J}^{0}\ (e)}\left( \Theta \right)  &\approx
&\left( I+\partial _{j}\otimes \mathbb{J}^{0}\right) W_{f}^{j\ (e)}\left(
\Theta \right) ,  \label{app} \\
W_{f}^{j+1+I\otimes \mathbb{J}^{0}\ (e)}\left( \Theta \right)  &\approx
&\left( I+\partial _{j}\otimes \left( I+\mathbb{J}^{0}\right) \right)
W_{f}^{j\ (e)}\left( \Theta \right) ,  \nonumber \\
W_{f}^{j+I\otimes \mathbb{J}^{0}\ (o)}\left( \Theta \right)  &\approx
&\left( I+\partial _{j}\otimes \left( I-\mathbb{J}^{0}\right) \right)
W_{f}^{j+1\ (o)}\left( \Theta \right) ,  \nonumber \\
W_{f}^{j+1+I\otimes \mathbb{J}^{0}\ (o)}\left( \Theta \right)  &\approx
&\left( I+\partial _{j}\otimes \mathbb{J}^{0}\right) W_{f}^{j+1\ (o)}.
\nonumber
\end{eqnarray}

Substituting~\eqref{pp}--\eqref{app} into the equations of motion
\begin{eqnarray}
i\partial _{t}W_{\rho }^{j\ (e)} &=&W_{H\rho }^{j\ (e)}\left( \Theta \right)
-W_{\rho H}^{j\ (e)}\left( \Theta \right) ,  \label{ep} \\
i\partial _{t}W_{\rho }^{j+1\ (o)} &=&W_{H\rho }^{j+1\ (o)}\left( \Theta
\right) -W_{\rho H}^{j+1\ (o)}\left( \Theta \right),  \label{ei}
\end{eqnarray}%
one obtains%
\begin{eqnarray*}
i\partial _{t}W_{\rho }^{j\ (e)} &\approx &2\left\{ W_{H}^{j\ (e)},W_{\rho
}^{j\ (e)}\right\} +2\left\{ W_{H}^{j+1\ (o)},W_{\rho }^{j+1\ (o)}\right\} ,
\\
i\partial _{t}W_{\rho }^{j+1\ (o)} &\approx &2\left\{ W_{H}^{j\ (e)},W_{\rho
}^{j+1\ (o)}\right\} +2\left\{ W_{H}^{j+1\ (o)},W_{\rho }^{j\ (e)}\right\} ,
\end{eqnarray*}%
where%
\begin{eqnarray}
\left\{ .,.\right\} &=&-\frac{\cot \theta }{j+1}\left( \partial _{\theta
}\otimes \partial _{\psi }-\partial _{\psi }\otimes \partial _{\theta
}\right)  \label{PB} \\
&&+\frac{1}{(j+1)\sin \theta }\left( \partial _{\theta }\otimes \partial
_{\phi }-\partial _{\phi }\otimes \partial _{\theta }\right) +\left(
\partial _{\psi }\otimes \partial _{j}-\partial _{j}\otimes \partial _{\psi
}\right) ,\nonumber
\end{eqnarray}%
are the Poisson brackets on $T^{\ast }\mathcal{S}_{2}$. Introducing $W^{\
\pm }\left( \Theta ,j\right) $ symbols according to~\eqref{W+} and \eqref{W-}%
, we immediately arrive at the Liouville form~\eqref{ee}.

\section{}\label{AppendixB}

The functions $A(j)$, $B(j)$ and $C(j)$ have the form
\begin{eqnarray*}
A(j) &=&\frac{1}{j+1}\sqrt{\left( \left( L_{1}+L_{2}+1\right) ^{2}-\left( 
\frac{j+1}{2}\right) ^{2}\right) \left( \left( \frac{j+1}{2}\right)
^{2}-\left( L_{1}-L_{2}\right) ^{2}\right) }, \\
B(j) &=&\frac{1}{\sqrt{j\left( j+2\right) }}\left( \frac{j}{2}\left( \frac{j%
}{2}+1\right) -L_{2}\left( L_{2}+1\right) +L_{1}\left( L_{1}+1\right)
\right) , \\
C(j) &=&\frac{1}{2}\left( \frac{j}{2}\left( \frac{j}{2}+1\right)
-L_{1}\left( L_{1}+1\right) -L_{2}\left( L_{2}+1\right) \right) .
\end{eqnarray*}%
The classical equations of motion corresponding to the Hamiltonian~\eqref{SS}
are%
\begin{eqnarray*}
\partial _{t}\phi  &=&\frac{2\lambda B(j)}{j+1}-\frac{2\lambda A(j)}{j+1}%
\cot \theta \cos \psi , \\
\partial _{t}\theta  &=&\frac{2\lambda A(j)}{j+1}\cos \theta \sin \psi , \\
\partial _{t}\psi  &=&\frac{2\lambda A(j)}{j+1}\frac{\cos ^{2}\theta }{\sin
\theta }\cos \psi +\frac{j+1}{2}+\lambda \left( \frac{j+1}{\sqrt{j(j+2)}}%
-2B(j)\frac{4j+2j^{2}+1}{\left( j+2\right) \left( j+1\right) }\right) \cos {%
\theta ,} \\
&&+\frac{2\lambda A(j)\left[ \left( j+1\right) ^{4}-16\left( \left(
L_{1}+L_{2}+1\right) \left( L_{1}-L_{2}\right) \right) ^{2}\right] \sin {%
\theta }\cos {\psi }}{\left( \left( j+1\right) ^{2}-4\left(
L_{1}-L_{2}\right) ^{2}\right) \left( \left( j+1\right) ^{2}-4\left(
L_{1}+L_{2}+1\right) ^{2}\right) \left( j+1\right) }{,} \\
\partial _{t}j &=&2\lambda A(j)\sin \theta \sin \psi .
\end{eqnarray*}

\end{paracol} 
\reftitle{References}



\end{document}